\documentclass[conference]{IEEEtran}
\usepackage{tikz}
\usetikzlibrary{shapes.geometric, arrows,decorations.pathreplacing}

\definecolor{red}{gray}{0.1}
\definecolor{green}{gray}{0.6}
\definecolor{blue}{gray}{0.3}
\definecolor{yellow}{gray}{0.6}
\definecolor{violet}{gray}{0.5}
\definecolor{orange}{gray}{0.9}

\tikzstyle{startstop} = [rectangle, rounded corners, minimum width=3cm, minimum height=0.75cm,text centered, draw=black, fill=red!30]
\tikzstyle{io} = [trapezium, trapezium left angle=70, trapezium right angle=110, minimum width=3cm, minimum height=0.75cm, text centered, draw=black, fill=blue!30]
\tikzstyle{process} = [rectangle, minimum width=3cm, minimum height=0.75cm, text centered, text width=3cm, draw=black, fill=orange!30]
\tikzstyle{decision} = [diamond, minimum width=3cm, minimum height=0.75cm, text centered, draw=black, fill=green!30]
\tikzstyle{arrow} = [thick,->,>=stealth]

\usepackage{cite}

\ifCLASSINFOpdf
  \graphicspath{{./figures/}{../jpeg/}}
  \DeclareGraphicsExtensions{.pdf,.jpeg,.png}
\else
\fi
\usepackage{amsmath}
\begin{document}
%
\title{StegIbiza: Steganography in Club Music\\ Implemented in Python}

\author{
\IEEEauthorblockN{Krzysztof Szczypiorski}
\IEEEauthorblockA{Warsaw University of Technology, Warsaw, Poland\\
Cryptomage SA, Wroclaw, Poland\\
Email: ksz@tele.pw.edu.pl}
\and
\IEEEauthorblockN{Wojciech Zydecki}
\IEEEauthorblockA{Warsaw University of Technology, Warsaw, Poland\\
Email: w.zydecki@gmail.com}}



%


\maketitle

\begin{abstract}
This paper introduces the implementation of steganography method called StegIbiza, which uses tempo modulation as hidden message carrier. With the use of Python scripting language, a bit string was encoded and decoded using WAV and MP3 files. Once the message was hidden into a music files, an internet radio was created to evaluate broadcast possibilities. No dedicated music or signal processing equipment was used in this StegIbiza implementation.
\end{abstract}


%
\IEEEpeerreviewmaketitle

\section{Introduction}
Steganography was used by human kind through out the history in many ways. We find first known examples in ancient Greece \cite{ancient_greece} and, as technology and culture progressed, new methods of steganography were developed. In the age of Internet major attention is paid to applications of steganography in image \cite{steg_image} and network \cite{steg_network} as it is currently best platform for covert communication, which pushes the construction of new image and network steganography methods. Recent years have shown usage growth of internet radio and music streaming services \cite{internet_radio_increase}, which provides plenty of room for new audio steganography methods.

In this paper, an implementation of new music steganography method called StegIbiza (Steganographic Ibiza) \cite{stegibiza} is introduced and used to share hidden information over an internet radio and other internet services. Implementation was done in Python programming language and was later used on pack of MP3 music files which were added to the internet radio station. The whole process was done using regular PC with out any special sound or signal processing equipment. The idea behind StegIbiza is to encode a message by increasing or decreasing the music tempo by small percent to keep it inaudible to humans. Music tempo is measured in beats per minute (bpm) and needs to be constant for the StegIbiza to work.

The paper is structured as follows: Section II briefly describes the art of steganography. Section III focuses on the idea of StegIbiza method and how it can be used. Section IV presents the limits which were discovered during the implementation process. Section V explains and presents the Python implementation of StegIbiza. Section VI shows usage of the implementation on different songs. Section VII presents possible broadcast sources like internet radio and other services. Section VIII concludes our work and describes our future efforts.  


 

\section{Art of Steganography}
Steganography is the practice of concealing information and communication. The word steganography combines two Greek words steganos, meaning "concealed" and graphein meaning "writing".

First documented examples of steganography usage are dated to 440 BC where a message was sent by tattooing it on slave head and concealed by regrown hair. As the time passed, many new steganography methods were used throughout the history like writing message on a back of a stamp placed on a postcard, or shrinking whole text to a size of a dot \cite{steg_history}. 

In nowadays the biggest communication channel for steganography is the internet and the all the services it provides. Major effort is put, into detection and prevention of any kind of concealed communication over the Internet \cite{steg_detecting}, in order to stop any terrorist groups from using steganagraphy to communicate.

The current state of the art in audio steganography is briefly presented in section II of  our previous work \cite{stegibiza} on StegIbiza.

\section{StegIbiza}
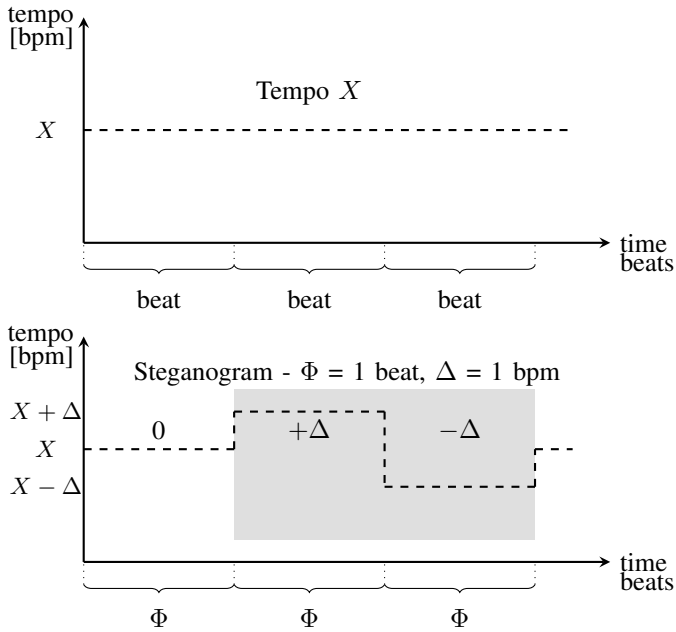
\begin{figure}[!t]
	\centering
	\begin{tikzpicture}
	\draw[arrow] (0,0) -- (7,0) node[anchor=west] {time} node[anchor=north west] {beats};
	\draw (3,2) node{{Tempo $X$}};
	
	\draw[arrow] (0,0) -- (0,3) node[anchor=east] {tempo} node[anchor=north east] {[bpm]};
	\draw (-0.5,1.5) node{{\small$X$}};
	
	\draw[thick,dashed] (0,1.5) -- (6.5,1.5);
	
	\draw[decorate,decoration={brace,amplitude=3pt,mirror}] (0,-0.3) -- (2,-0.3); 
	\draw[decorate,decoration={brace,amplitude=3pt,mirror}] (2,-0.3) -- (4,-0.3);
	\draw[decorate,decoration={brace,amplitude=3pt,mirror}] (4,-0.3) -- (6,-0.3);
	\node at (1,-0.75){beat};	
	\node at (3,-0.75){beat};
	\node at (5,-0.75){beat};
	\draw[dotted] (0,-0.3) -- (0,0);
	\draw[dotted] (2,-0.3) -- (2,0);
	\draw[dotted] (4,-0.3) -- (4,0);
	\draw[dotted] (6,-0.3) -- (6,0);
	\end{tikzpicture}
	\begin{tikzpicture}
		\fill[gray!25!white] (2,0.3) rectangle (6,2.3);
		
		\draw[arrow] (0,0) -- (7,0) node[anchor=west] {time} node[anchor=north west] {beats};
		\draw	(1,1.75) node{{0}}
		(3,1.75) node{{$+\Delta$}}
		(5,1.75) node{{$-\Delta$}};
		
		\draw (3.5,2.5) node{{Steganogram - $\Phi$ = 1 beat, $\Delta$ = 1 bpm }};
		
		\draw[arrow] (0,0) -- (0,3) node[anchor=east] {tempo} node[anchor=north east] {[bpm]};
		\draw (-0.5,1.5) node{{\small$X$}};
		\draw (-0.5,2) node{{\small$X+\Delta$}};
		\draw (-0.5,1) node{{\small$X-\Delta$}};

		\draw[thick,dashed] (0,1.5) -- (2,1.5);
		\draw[thick,dashed] (2,1.5) -- (2,2);
		\draw[thick,dashed] (2,2) -- (4,2);
		\draw[thick,dashed] (4,2) -- (4,1);
		\draw[thick,dashed] (4,1) -- (6,1);
		\draw[thick,dashed] (6,1) -- (6,1.5);
		\draw[thick,dashed] (6,1.5) -- (6.5,1.5);
		
		\draw[decorate,decoration={brace,amplitude=3pt,mirror}] (0,-0.3) -- (2,-0.3); 
		\draw[decorate,decoration={brace,amplitude=3pt,mirror}] (2,-0.3) -- (4,-0.3);
		\draw[decorate,decoration={brace,amplitude=3pt,mirror}] (4,-0.3) -- (6,-0.3);
		\node at (1,-0.75){$\Phi$};	
		\node at (3,-0.75){$\Phi$};
		\node at (5,-0.75){$\Phi$};
		\draw[dotted] (0,-0.3) -- (0,0);
		\draw[dotted] (2,-0.3) -- (2,0);
		\draw[dotted] (4,-0.3) -- (4,0);
		\draw[dotted] (6,-0.3) -- (6,0);
	\end{tikzpicture}
	\caption{Explanation of StegIbiza method ($\Phi$ = 1 beat, $\Delta$ = 1 bpm), hidden two symbols: "+" and "-"}
	\label{fig_stegibiza}
\end{figure}

Idea behind StegIbiza method is based on music tempo modulation. Using a song with constant tempo, a message is hidden by lowering and rising the tempo. In the initial StegIbiza implementation described in \cite{stegibiza} messages were encoded using adopted Morse code, where "dot was replaced with "+" and "dash was left as it is using "minus" symbol. The "plus" sing indicated a change +$\Delta$ for a period of $\Phi$ beats and the "minus" meant a change of -$\Delta$ for a period of $\Phi$ beats. Original tempo was marked as 0 and was taken as a reference. An example usage is presented in Figure \ref{fig_stegibiza}.

StegIbiza was evaluated on a group of testers in order to estimate the values of $\Phi$ and $\Delta$, that make tempo change inaudible for humans. In the evaluation 5 different songs were played to 20 testers, aged between 25 and 45 and without any hearing impairments. Tempo was increased by different values with constant $\Phi = 1$ beat. Seven of the testers had professional music background (graduated from musical school) and three of them worked as professional musicians. The evaluation has shown that the 1\% change in tempo was indiscoverable for anyone from the testing group. Changes above 1\% but below 2\% were noticeable only for professional musicians. With changes above 2\% but below 3\% around 50\% (9) of testers noticed the difference. With a change of 3\% and above all testers have detected the StegIbiza.

\section{Implementation Revelations}
Before implementing StegIbiza in Python, a research was made in order to find best solution and tools for the task. Most important part of the StegIbiza implementation was accurate tempo measurement and error free message decoding. For this three libraries were tested:
\begin{itemize}
	\item Aubio \cite{aubio}: free and opensource library written in C to label music and sounds.
	\item SoundTouch \cite{soundtouch}: opensource audio processing library for modifying tempo, pitch and playback rates of audio steams and files.
	\item madmom \cite{madmomIEEE}: audio signal processing library written in Python.
\end{itemize}
In the test, a song was divided into parts where $\Phi\in\{10,15,20\}$ seconds. Tempo of each part was measured and compared with the original one. Finally "madmom" was chosen due to providing best method of measuring tempo value in a song file. In comparison to other libraries, "madmom" had a constant error value, while measurements with "Aubio" and "SoundTouch" presented errors within $+/-1\%$ range. In addition to "madmom" for tempo reading, "SoundTouch" library is used for tempo modulation and brought to Python using a wrapper "pysoundtouch".

\begin{figure}[!t]
	\centering
	\includegraphics[width=3.5in]{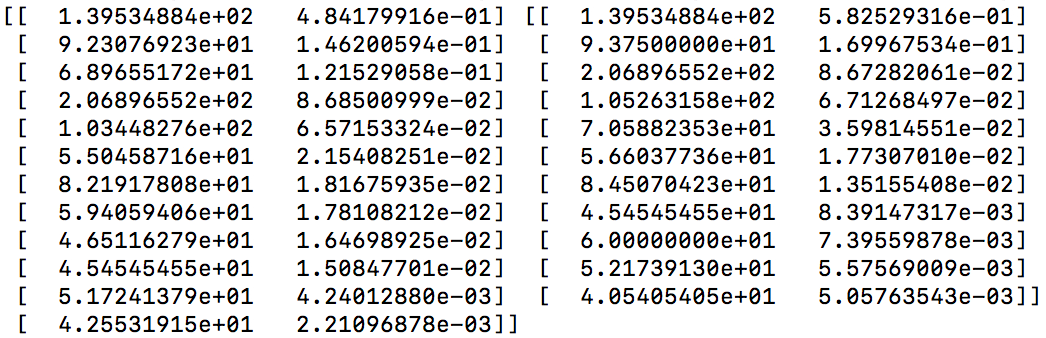}
	\caption{Example of madmom measurement. Array on the left presents the unchanged tempo of song starting from 0 to 10 seconds. The right side presents the tempo increased by 1\% starting from 10 to 20 seconds.}
	\label{fig_madmom}
\end{figure}

When it came to measurement accuracy, none of the libraries could correctly detect tempo change that was lower than 2,75\%. At first this seemed like an big issue, because the goal was to read $+/-1\%$  tempo changes, so that StegIbiza stays inaudible. However in the end there is no need for accurate change measurement, as long as the results allow us to determine if the tempo was increased or decreased. Measurement results from "madmom" are presented as an array of BPM values and beat strength. Example measurement results of two files is presented in Figure \ref{fig_madmom}. First array on the left contains results of unchanged tempo measurement of first 10 seconds of the song. Array on the right presents tempo measurement results from 10 till 20 seconds of the same song with its tempo increased by 1\%. When we compare the results we can notice small changes in some values and identify the increase.

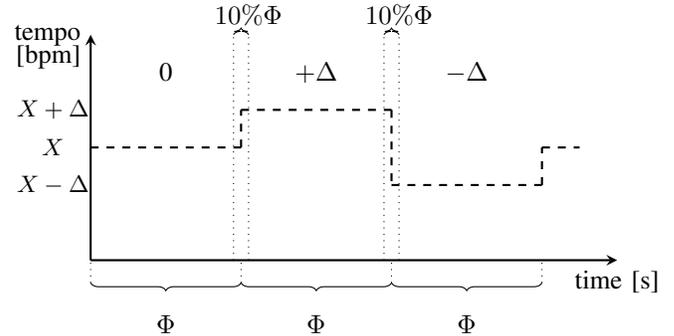
\begin{figure}[!t]
	\centering
	\begin{tikzpicture}
	\draw[arrow] (0,0) -- (7,0) node[anchor=north] {time [s]};
	\draw	(1,2.5) node{{0}}
	(3,2.5) node{{$+\Delta$}}
	(5,2.5) node{{$-\Delta$}};
	
	\draw[arrow] (0,0) -- (0,3) node[anchor=east] {tempo} node[anchor=north east] {[bpm]};
	\draw (-0.5,1.5) node{{\small$X$}};
	\draw (-0.5,2) node{{\small$X+\Delta$}};
	\draw (-0.5,1) node{{\small$X-\Delta$}};

	\draw[dotted] (1.9,0) -- (1.9,3);
	\draw[dotted] (2.1,0) -- (2.1,3);
	\draw[decorate,decoration={brace}] (1.9,3) -- (2.1,3) node[anchor=south] {$10\%\Phi$}; 
	\draw[dotted] (3.9,0) -- (3.9,3);
	\draw[dotted] (4.1,0) -- (4.1,3);
	\draw[decorate,decoration={brace}] (3.9,3) -- (4.1,3) node[anchor=south] {$10\%\Phi$}; 
	
	\draw[thick,dashed] (0,1.5) -- (2,1.5);
	\draw[thick,dashed] (2,1.5) -- (2,2);
	\draw[thick,dashed] (2,2) -- (4,2);
	\draw[thick,dashed] (4,2) -- (4,1);
	\draw[thick,dashed] (4,1) -- (6,1);
	\draw[thick,dashed] (6,1) -- (6,1.5);
	\draw[thick,dashed] (6,1.5) -- (6.5,1.5);
	
	\draw[decorate,decoration={brace,amplitude=3pt,mirror}] (0,-0.3) -- (2,-0.3); 
	\draw[decorate,decoration={brace,amplitude=3pt,mirror}] (2,-0.3) -- (4,-0.3);
	\draw[decorate,decoration={brace,amplitude=3pt,mirror}] (4,-0.3) -- (6,-0.3);
	\node at (1,-0.85){$\Phi$};	
	\node at (3,-0.85){$\Phi$};
	\node at (5,-0.85){$\Phi$};
	\draw[dotted] (0,-0.3) -- (0,0);
	\draw[dotted] (2,-0.3) -- (2,0);
	\draw[dotted] (4,-0.3) -- (4,0);
	\draw[dotted] (6,-0.3) -- (6,0);
	\end{tikzpicture}
	\caption{Explanation of StegIbiza implementation ($\Phi$ - time in seconds, $\Delta$ - 1\% of default tempo), hidden two symbols "+" and "-"}
	\label{fig_stegibiza_imp}
\end{figure}

The process of implementation has also shown that lowest possible value of $\Phi = 10$ seconds. It turned out that for decent tempo measurement result, one needs at least 9 seconds of a song, which would make constant $\Phi = 9$ seconds, however when we increase or decrease the tempo, we also change the length of the $\Phi$ parts. When reading the encoded value with $\Phi = 9$ we might read a small part of the previous or next song part with different tempo. This will lead to an error and will most likely end in reading wrong value. To avoid that we cut 5\% of the beginning and end of each song part while reading, hence $\Phi = 10 =  9 + 10*5\% + 10*5\%$. The implementation is presented in Figure \ref{fig_stegibiza_imp}. 

For  $\Phi = 10$, with average song time of 3 minutes we can only hide 17 bits - $3*60/10 - 1$ (we subtract 1 bit as it is used for original tempo reference). By the cost of songs data capacity, we can increase the $\Phi$ value to 15 or 20 seconds which will lower the probability of an error and StegIbiza detection.

\section{StegIbiza Python Implementation}
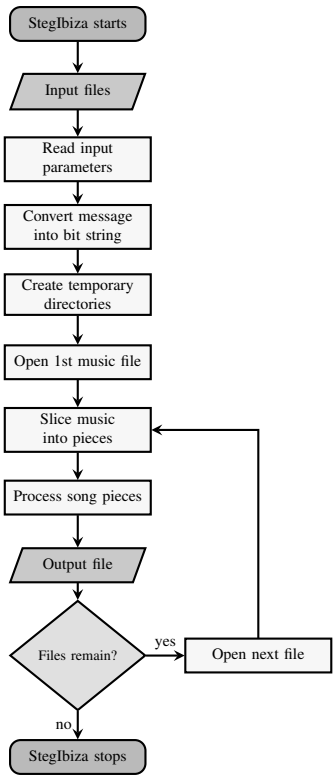
\begin{figure}[!t]
	\centering
	\begin{tikzpicture}[node distance=1.5cm, thick, scale=0.6, every node/.style={scale=0.6}]
		\node (start)			[startstop]              {StegIbiza starts};
		\node (in1)				[io, below of=start] 	{Input files};
		\node (read)     [process, below of=in1]          {Read input parameters};
		\node (processMessage)      [process, below of=read]   {Convert message into bit string};
		\node (tmpDirs)      [process, below of=processMessage]   {Create temporary directories};
		\node (openFile)      [process, below of=tmpDirs]   {Open 1st music file};
		\node (sliceFile)      [process, below of=openFile]   {Slice music into pieces};
		\node (process)      [process, below of=sliceFile]   {Process song pieces};
		\node (output)				[io, below of=process] 	{Output file};
		\node (dec) 		[decision, below of=output, yshift=-0.5cm] {\small{Files remain?}};
		\node (nextfile)      [process, right of=dec, xshift=2.5cm]   {Open next file};
		\node (stop)			[startstop, below of=dec, yshift=-0.75cm]              {StegIbiza stops};
		{onPause()};
		\draw [arrow] (start) -- (in1);
		\draw [arrow] (in1) -- (read);
		\draw [arrow] (read) -- (processMessage);
		\draw [arrow] (processMessage) -- (tmpDirs);
		\draw [arrow] (tmpDirs) -- (openFile);
		\draw [arrow] (openFile) -- (sliceFile);
		\draw [arrow] (sliceFile) -- (process);
		\draw [arrow] (process) -- (output);
		\draw [arrow] (output) -- (dec);
		\draw [arrow] (dec) -- node[anchor=east] {no} (stop);
		\draw [arrow] (dec) -- node[anchor=south] {yes} (nextfile);
		\draw [arrow] (nextfile) |- (sliceFile);
	\end{tikzpicture}
	\caption{Flowchart of StegIbiza implementation.}
	\label{flowchart_stegibiza}
\end{figure}

The implementation takes WAV or MP3 type music file and slices it into $\Phi=10$ second parts by default. First part remains untouched as it is used as a reference for original tempo value. The tempo of the remaining parts is modified accordingly to bits of the hidden message. Time of the last part of the song file is always lower than $\Phi$ therefor it remains untouched as it is very likely to produce errors when decoding a message. After the encoding all parts are put back together into one file. For long messages it is possible to provide several music files for increased storage capacity. Figure \ref{flowchart_stegibiza} presents simplified flowchart of StegIbiza python implementation.

When decoding a message, the file is split into $\Phi - 10\%\Phi$ parts. Here we cut the beginning and end by $5\%$ of $\Phi$ of each part in order to eliminate overlapping with neighboring parts. Once the file is divided, we start reading the message by comparing the tempo of each file with the first one and qualify it if has increased or decreased tempo. Same as before, time of the last part of the song file is always lower than $\Phi$ therefor it remains untouched.

In order to classify the measured sample tempo as +1\% or -1\% a simple classifier is used. In order to obtain attributes of the sample, each tempo value from the reference song slice is compared with each tempo value of the analyzed sample and saved as difference in \%. Then all obtained attributes with mean value larger than 4\% are discarded. Remaining features are summed and classified as +1\% if the sum is larger than 0 or as -1\% otherwise. 

\section{Implementation Results}
The implementation was tested on 10 different songs obtained from "Free Music Archive" \cite{freemusicarchive}. The test results are presented in Table \ref{table_results}. The bit string used in test is composed of 21 bits, therefor requires $21*10s = 3min \& 30s$ for the message to be fully encoded. Some songs used in the test were shorter than the time required for the whole message to be encoded and the missing bits are represented as "x" in the table.

Songs numbered from 1 till 6 are categorized as electronic music and they have constant tempo. Form the results we see that songs from 1 to 5 are error free. Song number 6 presents interesting result as it is inverted from the original bit string. This was caused by 5 seconds silence in the beginning of the song, which led to bad tempo reference sample and caused reverted results.

Songs numbered 7 and 8 are classified as instrumental indie-rock. Those songs produce some errors when decoding hidden message due to small tempo changes in some parts.

Songs 9 and 10 do not have constant tempo and produce large number of errors. In this case there is no way to use StegIbiza in current form. The only possibility here would be to compare each part of the song with equivalent part from the original song version.

\begin{table*}[!t]
\renewcommand{\arraystretch}{1.3}
\caption{Results of message encoding and decoding using StegIbiza implementation in Pyhton}
\label{table_results}
\centering
\begin{tabular}{|c|c|c|c|c|}
\hline
\textbf{\#} & \textbf{Artist and Song Name} & \textbf{Original Duration} &  \textbf{Decoded Bit String} & \textbf{\#Errors}\\
\hline
\hline
\textbf{0} & \textbf{Original Bit String} & 3:30  & 1 1 0 1 1 0 1 1 1 1 0 0 1 1 1 1 0 0 1 1 1 & X\\
\hline
\hline
1 & Broke For Free - Night Owl \cite{night_owl} & 3:14  & 1 1 0 1 1 0 1 1 1 1 0 0 1 1 1 1 0 x x x x & 0\\
\hline
2 & FLASERS - Amsterdam \cite{amsterdam} & 3:42 & 1 1 0 1 1 0 1 1 1 1 0 0 1 1 1 1 0 0 1 1 x & 0\\
\hline
3 & Lobo Loco - Muscle Body Man (ID 534) \cite{muscle_body_man} & 5:57 & 1 1 0 1 1 0 1 1 1 1 0 0 1 1 1 1 0 0 1 1 1 & 0\\
\hline
4 & Tours - Enthusiast \cite{enthusiast} & 2:51 & 1 1 0 1 1 0 1 1 1 1 0 0 1 1 1 x x x x x x & 0\\
\hline
5 & Black Ant - Fater Lee \cite{fater_lee}& 2:23 & 1 1 0 1 1 0 1 1 1 1 0 0 x x x x x x x x x & 0\\
\hline
6 & Candlegravity - Tomie\'s Bubbles \cite{tomies_bubbles} & 6:21 & \textbf{0 0 1 0 0 1 0 0 0 0 1 1 0 0 0 0 1 1 0 0 0} & 21\\
\hline
7 & Jahzzar - Siesta \cite{siesta} & 2:20 & 1 1 0 1 1 0 1 1 1 \textbf{0 1 1} x x x x x x x x x & 3\\
\hline
8 & Gillicuddy - Springish \cite{springish} & 2:23 & 1 1 \textbf{1} 1 \textbf{0} 1 1 1 1 \textbf{0} 0 x x x x x x x x x x & 3\\
\hline
9 & Podington Bear - Pick Up The Tempo \cite{pick_up_the_tempo} & 2:40 & 1 1 0 1 \textbf{0} 1 \textbf{0 0 0 0 1 1 0 0} x x x x x x x & 9\\
\hline
10 & The Kyoto Connection - Hachiko (The Faithtful Dog) \cite{hachiko} & 3:05 & 1 1 0 1 1 0 \textbf{0 0 0} 1 0 \textbf{1} 1 1 \textbf{0 1} x x x x x & 6\\
\hline
\end{tabular}
\end{table*}

\section{StegIbiza Broadcast Applications}
The easiest approach to broadcast information using StegIbiza is to simply upload the music files to a public website as a playlist and allow the receiver to download them and decode the message. Any other random person to download the playlist or any song in it, would not recognize any tempo changes by simply listening to it.

Another simple way is to put the songs with hidden message on popular video service like YouTube \cite{youtube}. Then the receiver can extract song file from the video and decode the hidden message. 

Slightly harder solution is to setup an internet radio using SouthCast \cite{southcast} or other tools. Then we can setup automatic broadcast of previously prepared playlist with a hidden message. Each song should be separated by a proper separator like few seconds of silence or some other sound mark that will indicate new song. Later to extract message from each song, the stream can be easily recorded and split into song files using the separators.

There is always a possibility of using StegIbiza in regular radio, however messages are most likely to be unreadable due to bad quality recordings caused by bad reception and interferences.

Besides hiding messages in music, idea of tempo changing from StegIbiza can be used to control variety of equipment in clubs or music festivals. Lighting, screens and many others devices can be wirelessly controlled by reacting to music tempo changes.

\section{Conclusion}
This article presented StegIbiza implementation and application on different music tracks. It showed that StegIbiza can be applied using regular personal computer without any special music or signal processing hardware. Implementation has also revealed its constrains in terms of tempo change limits and song capacity for data. The implementation was tested on 10 different songs and proven to work with 1\% tempo change, with 1 bit saved in each 10 seconds of the song. Article has also presented simple ways to secretly share information hidden by StegIbiza using internet services.

In this implementation there is a lot of space for improvements and the future work will focus on lowering the tempo change to lowest possible values and increasing songs data capacity by lowering the minimum time value needed to read the tempo.


\section*{Acknowledgment}
The authors would like to thank all the people that helped in StegIbiza research or developed tools that made this implementation possible.

This is an independent publication and has not been authorized, sponsored, or otherwise approved by Google, SouthCast or Free Music Archive.



%

\end{document}